# A Performance Metric for Discrete-Time Chaos-Based Truly Random Number Generators

Ahmad Beirami, Hamid Nejati, and Yehia Massoud
Electrical and Computer Engineering Department, Rice University, Houston TX 77005
(massoud@rice.edu)

*Abstract*—In this paper, we develop an information entropy-based metric that represents the statistical quality of the generated binary sequence in Truly Random Number Generators (TRNG). The metric can be used for the design and optimization of the TRNG circuits as well as the development of efficient post-processing units for recovering the degraded statistical characteristics of the signal due to process variations.

## I. Introduction

A truly random number generator (TRNG) is a binary generator that is capable of producing independent and unbiased bits in a nondeterministic and irreproducible process [1]. A TRNG is a critical part of cryptographic systems, where it provides the necessary uncertainty or entropy in order to secure the message transmission. Recently, discrete-time chaotic systems have been significantly used in the design of high speed TRNGs. Chaos is the non-predictive behavior of certain nonlinear dynamical systems that happens due to the high sensitivity to the initial state. Any small perturbation can grow exponentially in the system and can result in a non-predictive chaotic behavior [2].

A discrete-time chaotic map, formed by the iteration of the signal in a simple nonlinear map function, can be used for the generation of truly random numbers. Piecewise linear input-output (I/O) characteristic curves have been extensively used as the map function for the generation of a binary output sequence according to the state of the piecewise linear function [1]. The capability of integration, high speed, and the high quality of generated bits due to the simple discrete-time nature of the system make the discrete-time chaotic maps very good candidates for the generation of random numbers.

In practice, the process variations in the circuit implementation usually lead to a degradation in the statistical characteristics of the generated binary sequence. This necessitates the utilization of a post-processing unit to recover the statistical characteristics of the bit sequence. Von-Neumann [3], SHA-1 [4], and SHA-2 [5] are the most widely used post-processing algorithms. All of the post-processing algorithms should trade the bit-rate off with true randomness of the generated binary sequence. However, these algorithms often post-process the data *blindly*, i.e., without any knowledge of the non-ideal nature of the binary sequence. The development of methods that can predict the statistical characteristics of the binary sequence can enable the design of efficient post-processing units that can increase the efficiency of the TRNG.

The true randomness of the generated binary sequence is traditionally validated against NIST 800-22 randomness test suite [6]. NIST 800-22 is a pass/fail test that takes a large binary sequence as input and can only prove that a sequence is not random. This randomness test can not be used for the evaluation of the *true* randomness of the binary sequence. NIST 800-22 randomness test suite also does not provide the designer with much information regarding the efficiency of the design in order to optimize the performance of the TRNG circuits.

In this paper, we propose a metric for the performance evaluation of the discrete-time chaos-based TRNG circuits. The information entropy of the generated binary sequence can represent the quality of the statistical characteristics of the generated bits. We develop a methodology to find the entropy of the generated bits according to the discrete-time chaotic map.

## II. Discrete-Time Chaotic Sources: Fundamentals

Discrete-Time chaotic maps are formed by the iteration of the output signal in a transformation function $M(x) : (0, 1) \to (0, 1)$ in a positive feedback loop as given by

$$x_{n+1} = M(x_n) = M^n(x_0). \tag{1}$$

In this equation, $n$ represents the time step, $x_0$ is the initial output signal (initial state) of the system, and $x_n$ is the output signal at time step $n$. For example, for the sample map function shown in Figure 1(a), the transformation function is given by

$$x_{n+1} = M(x_n) = \frac{3\sqrt{3}}{2} x_n (1 - x_n^2). \tag{2}$$

As shown in Figure 1(a), the input and output ranges are both $(0, 1)$. Note that $(0, 1)$ is an arbitrary choice without loss of generality since any range can be scaled to $(0, 1)$.

For a discrete-time chaotic map, a density distribution function $f_n(x)$ is defined as a function of the input parameter $x$ and the time step $n$. $f_n(x)$ demonstrates the density of possible states in the system at time step $n$ (after $n$ iterations). A higher $f_n(x)$ demonstrates a higher probability of being at $x$ at time $n$. Since the output of the system at time $n + 1$ is only a function of the output of the system at time $n$, $f_{n+1}(x)$ can be obtained recursively from $f_n(x)$ using the Frobenius-Perron operator $P$, as given by [7]

$$f_{n+1}(x) = P f_n(x). \tag{3}$$

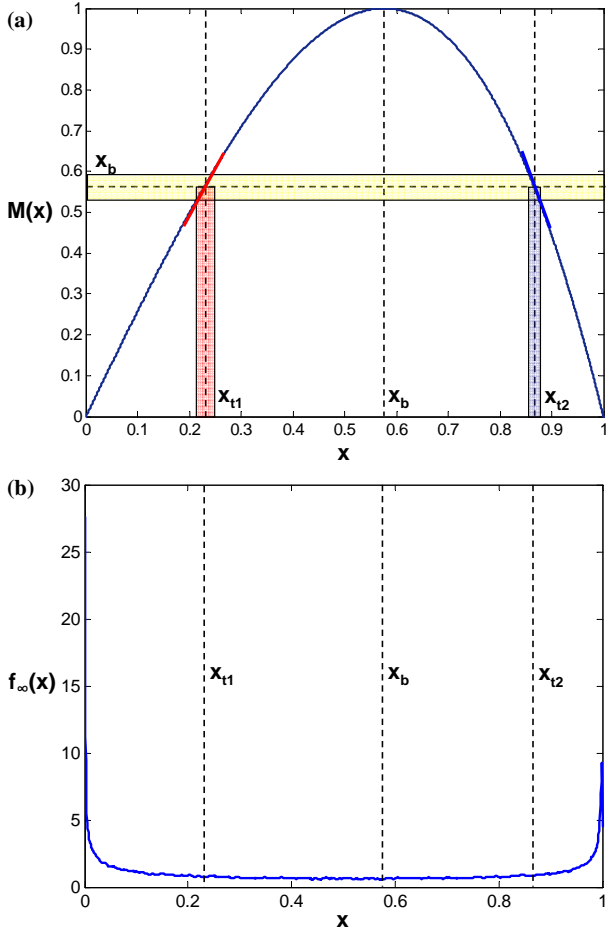

Fig. 1. (a) The example map. (b) The density distribution of the map.

For example, in the sample map shown in Figure 1(a), the density of states in $x_b$ at time $n+1$, $f_{n+1}(x_b)$ is obtained as follows. $x_b$ is the output value of the system for input values $x_{t1}$ and $x_{t2}$. The transformation of the density of states from $x_{t1(2)}$ to $x_b$ is inversely proportional to the slope of the map as shown in Figure 1(a). Therefore, $f_{n+1}(x_b)$ can be given by

$$f_{n+1}(x_b) = \frac{1}{|M'(x_{t1})|}f_n(x_{t1}) + \frac{1}{|M'(x_{t2})|}f_n(x_{t2}), \quad (4)$$

where $M'(x)$ is the derivative of the map function $M(x)$ at point $x$.

Under certain conditions, which are usually satisfied for TRNG maps, the density distribution asymptotically converges to a stable distribution $f_\infty(x)$ as $n \to \infty$, regardless of the initial value of the system [8]. In this case, $f_\infty(x)$ can be given by solving

$$f_\infty(x) = Pf_\infty(x). \quad (5)$$

The analytical solution of the asymptotic density distribution is not straight-forward for an arbitrary map function due to the explicit definition of the Frobenius-Perron operator $P$. Therefore, we develop a simulation-based method to obtain the asymptotic density distribution of a discrete-time chaotic map. The ergodic nature of the problem enables the finding of the asymptotic density distribution by the iteration of an arbitrary initial point in the map function and following the output trajectory to find the state density distribution. However, this method needs to be carefully implemented since the digital nature of the computer can lead to periodic patterns in the output. We divide $(0,1)$ to $L$ equal parts that are represented by $t_0, ..., t_{L-1}$, where $t_i = i/L$. We then store the values of the map function for the $L$ selected values $M(t_0), ..., M(t_{L-1})$. The initial value of the map $\tilde{x}_0$ is randomly selected from $t_0, ..., t_{L-1}$. We define the iterative relation between input and output as

$$\tilde{x}_{n+1} = \frac{1}{L}floor[L \times M(\tilde{x}_n) + a_n], \quad (6)$$

where $a_n$ is a random variable with uniform distribution in $(-1, 1)$, and $\tilde{x}_n$ is the digital output value at time $n$. This I/O relation ensures that $\tilde{x}_{n+1} \in t_1, ..., t_L$, which is the necessary condition for the iteration of the output in the system. The random variable $a_n$ is added to prevent the periodic patterns that can appear due to the digitization of the continuous map function. If $L$ is chosen very large, $a_n$ will look like a noise in the order of $1/L$.

The first several iterations of the system need to be discarded in order to ensure that the output value has reached the asymptotic density distribution. In the next $K$ iterations, we count the number of visits to each point $t_i$, and call it $Num(t_i)$. The density of states at point $t_i$ is given by

$$\tilde{f}_n(t_i) = \frac{L}{K}Num(t_i). \quad (7)$$

Note that increasing $L$ and $K$ will increase the resolution of the response, while requiring more simulation time. Moreover, since $Num(t_i)$ is an integer function, the resolution of the density function is limited to $L/K$. Therefore, it is necessary that $K >> L$ for an acceptable resolution. In Figure 1(b), the asymptotic density distribution of the sample map $M(x)$ has been obtained by using the developed method. In this map, the very low slope of the map near $x = x_b$, where $M(x_b) = 1$, leads to high density of states near $x = 1$. This is due to the transformation of a large span in the neighborhood of $x_b$ into a much smaller span in the neighborhood of 1. The high density of states near $x = 0$ is due to the fact that the neighborhood of $x = 1$ is transformed to the neighborhood of $x = 0$. Therefore, a high density near $x = 1$ results in a high density near $x = 0$.

### III. STATISTICAL CHARACTERISTICS OF THE GENERATED BINARY SEQUENCE

In order to generate a binary sequence from a discrete-time chaotic map, $(0,1)$ needs to be partitioned to two bit-generation partitions $S(0)$ and $S(1)$, which correspond to the bits 0 and 1, respectively. Any of the partitions $S(0)$ or $S(1)$ can be in general non-connected. For example, in the sample map shown in Figure 1, we assume that $S(0) = (0, x_b)$, and $S(1) = (x_b, 1)$, where $x_b = 1/\sqrt{3}$ is the abscissa of the maximum point.

Output value $x$ corresponds to 0 (or 1), if and only if $x \in S(0)$ (or $x \in S(1)$). Next, we develop a methodology for the calculation of the information entropy of the output sequence, as a metric for the evaluation of the statistical characteristics of the generated binary sequence.

A sequence of $N$ bits, called an $N$-bit sequence, has $2^N$ possible cases according to the state of each bit. We find the partitions that can lead to each of the $2^N$ cases. $S_N(i_1,...,i_N)$ is defined as the partition that includes all the initial state points $x$ that will generate the $N$-bit sequence of $(i_1,...,i_N)$, where $i_1,...,i_N \in \{0,1\}$. For example, $S_3(0,0,1)$ is a partition of $(0,1)$ that includes all the initial state points that will lead to the generation of the sequence $(0,0,1)$ in the three iterations. Note that $S_1 = S$ corresponds to the bit-generation partitions $S_1(0) = S(0)$ and $S_1(1) = S(1)$, which are defined by the bit-generation circuit design. All $S_N$ for $N \geq 2$ can be found based on $S$ and the map function $M(x)$. $S_N$ is related to $S$ through the following relation

$$x_1 \in S_N(i_1,...,i_N) \Leftrightarrow x_j \in S(i_j), \ j=1,...,N, \quad (8)$$

where $i_j \in \{0,1\}$. We find the $2^{N+1}$ partitions for an $(N+1)$-bit iteratively based on the $2^N$ partitions for an $N$-bit sequence. Assume that $S_N(i_1,...,i_N)$ is known for $i_1,...,i_N \in \{0,1\}$. According to the equation (8), it can be shown that

$$x_1 \in S_N(i_1,...,i_N) \Leftrightarrow x_2 \in S_{N-1}(i_2,...,i_N), \quad (9)$$

$$x_1 \in S_{N+1}(i_1,...,i_N,i_{N+1}) \Leftrightarrow x_2 \in S_N(i_2,...,i_{N+1}). \quad (10)$$

Therefore, if $x_1 \in S_N(i_1,...,i_N)$, then either $x_2 \in S_N(i_2,...,i_N,0)$ or $x_2 \in S_N(i_2,...,i_N,1)$. For each initial point $x_1$, we find the output $x_2$. Then, we find the corresponding partition of $x_2$ among all $S_N$ since $2^N$ partitions of $S_N$ are known. Equation (10) shows how this can lead to finding the corresponding partition of $x_1$ among all $2^{N+1}$ partitions of $S_{N+1}$.

In the next step, we find the probability of the occurrence of each of the $2^N$ cases in an $N$-bit sequence. It is reasonable to assume that the output value follows the asymptotic density distribution since there is no access to the output value in the output binary sequence that can perturb the distribution. The occurrence probability of the case $(i_1,...,i_N)$ is called $P_N(i_1,...,i_N)$, and is given by the integration of the asymptotic density distribution over the corresponding partition $S_N(i_1,...,i_N)$, as given by

$$P_N(i_1,...,i_N) = \int_{S_N(i_1,...,i_N)} f_\infty(x)dx, \quad (11)$$

where $i_1,...,i_N \in \{0,1\}$. For the case of $N=1$, the probabilities $P_1(0) = P(0)$ and $P_1(1) = P(1)$ demonstrate the steady state probability of occurrence of 0 and 1 in the system. The bias, $b$, in the binary sequence is defined by

$$b = |P(0) - \frac{1}{2}| = |P(1) - \frac{1}{2}|. \quad (12)$$

The bias is the most important metric for the randomness of binary sequences.

For example, in the sample map shown in Figure 1(a), we have $S_1(0) = (0,x_b)$ and $S_1(1) = (x_b,1)$. Therefore, $P(0)$ and $P(1)$ are given by

$$P(0) = \int_0^{x_b} f_\infty(x)dx,$$
$$P(1) = \int_{x_b}^1 f_\infty(x)dx, \quad (13)$$

where $P(0) = 0.57$ and $P(1) = 0.43$. The partitions $S_2(0,0), S_2(0,1), S_2(1,0), S_2(1,1)$ are also demonstrated in Figure 1(a), and are given by

$$S_2(0,0) = (0,x_{t1}), \quad S_2(0,1) = (x_{t1},x_b),$$
$$S_2(1,0) = (x_{t2},1), \quad S_2(1,1) = (x_b,x_{t2}), \quad (14)$$

where $x_{t1} = 0.24$ and $x_{t2} = 0.86$. The probability of occurrence of each of the 2-bit sequences can be calculated using (11) by integrating the asymptotic density distribution over the corresponding partitions, and is given by

$$P_2(i_1,i_2) = \int_{S_2(i_1,i_2)} f_\infty(x)dx \quad (15)$$

for $i_1,i_2 = \{0,1\}$. We have $P_2(0,0) = 0.35$, $P_2(0,1) = 0.22$, $P_2(1,0) = 0.23$ and $P_2(1,1) = 0.20$ for the sample map.

## IV. INFORMATION ENTROPY AS A PERFORMANCE METRIC

In this section, we calculate the information entropy as the performance evaluation metric. The information entropy $H$ for a single bit $i$ is defined as [9]

$$H = -P(i=0)\log_2 P(i=0) - P(i=1)\log_2 P(i=1). \quad (16)$$

It can be shown that $0 \leq H \leq 1$. This equation implies that the information entropy of the bit-generation process is equal to $H$ bits. $H$ takes its maximum value of 1 bit if $P(0) = P(1) = 1/2$, i.e., for this case the number of uncertain bits is equal to 1. $H$ is equal to 0 if $P(0) = 0$ or $P(0) = 1$. These two cases stand for the deterministic bit-generation process and there is no uncertainty or entropy in the process. For $0 < P(0) < 1/2$, the information entropy and the uncertainty will increase from the minimum of 0 bit to the maximum of 1 bit. In this case, the likelihood of the occurrence of a 1 is more than the likelihood of the occurrence of a 0. This will lead to a decrease in the information entropy.

The information entropy $H_N$ of an $N$-bit sequence is defined according to the probability of occurrence of each of the $2^N$ sequences, as given by

$$H_N = - \sum_{i_1,...,i_N \in \{0,1\}} P_N(i_1,...,i_N) \log P_N(i_1,...,i_N). \quad (17)$$

It can be shown that $0 < H_N < N$, i.e., the information entropy of an $N$-bit sequence is at most equal to $N$ bits. In other words, there are at most $N$ uncertain bits in a sequence of $N$ bits.

We define $h_N$ as the entropy of the $N$-th generated bit, as given by

$$h_N = H_N - H_{N-1}. \quad (18)$$

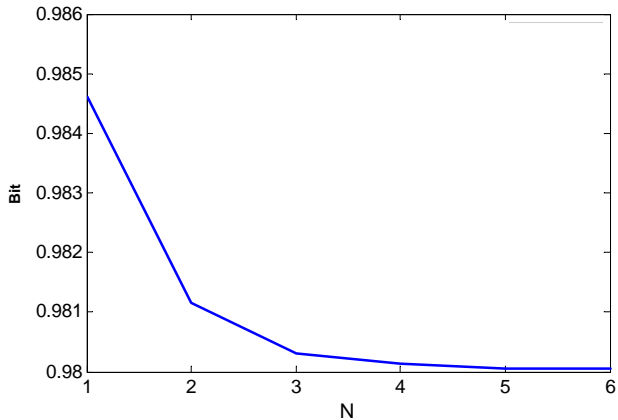

Fig. 2. $h_N$: the information entropy of the $N$-th generated bit.

$h_N$ is the information entropy of the generation of a bit, where the previous $N-1$ generated bits are known. The relation between $H_N$ and $h_N$ can be written as

$$H_N = \sum_{k=1}^{N} h_k. \quad (19)$$

$h_N$ is a monotonically decreasing sequence of $N$ since the knowledge of $N+1$ previous bits can decrease the entropy of the bit-generation more than the knowledge of $N$ previous bits. It can also be shown that $h_N$ converges to a fixed value $h$ as $N \to \infty$, given by

$$h = \lim_{N \to \infty} h_N. \quad (20)$$

We propose $h$ as a metric for the performance of truly random number generators. We can approximate $h$ with $h_N$ and since $h_N$ is monotonically decreasing with $N$, we can decrease the approximation error as much as necessary by increasing $N$.

We used the developed method to calculate the information entropy of the bits for the sample map. In Figure 2, $h_N$ calculated by the developed method is demonstrated. As expected, both sequences are monotonically decreasing and both converge to the same point. An asymptotic entropy of more than $0.98$ is achieved, which demonstrates a good quality of the generated bit sequence.

$h$ can also be utilized for design of an efficient post-processing unit. The degraded information entropy of the generated bit sequence, makes it necessary to decrease the bit-rate $R$ of the sequence in order to obtain truly random bits. This is due to the fact that any deterministic processing performed on the sequence can not compensate for the lost information entropy. At least, the bit-rate $R$ of the sequence has to be decreased by a factor of $h$ to ensure that the information entropy of the resulting sequence can meet the threshold for being truly random, as given by

$$R_d = \frac{1}{h} R, \quad (21)$$

where $R_d$ is the decreased bit-rate. Note that special algorithms must be used to ensure that the generated binary sequence is truly random.

## V. CONCLUSION

In this paper, we developed a methodology to calculate the information entropy of the bits generated from a discrete-time chaotic map. Information entropy is proposed as a metric for performance evaluation of the discrete-time chaos-based TRNGs. The developed metric enables the efficient design and optimization of TRNG circuits. The information entropy metric can also be used for the design of efficient post-processing units for the truly random number generators.